\documentclass[%
 preprint, 
 amsmath,amssymb,
 aps, physrev,
]{revtex4-2}

\providecommand{\texorpdfstring}[2]{#1}

\usepackage{graphicx}
\usepackage{dcolumn}
\usepackage{bm}
\usepackage{hyperref}
\usepackage{multirow}
\usepackage{dcolumn}
\usepackage{float}

\begin{document}

\preprint{APS/123-QED}

\title{$f_0(980)$ production from $K\bar{K}$ coalescence in pp collisions at $\sqrt{s}=5.02$ TeV within UrQMD}

\author{Phacharatouch Chaimongkon}
\author{Krittaporn Anukulkitch}
\author{Pornrad Srisawad}
	\email{Contact corresponding author : pornrads@nu.ac.th}
\affiliation{
 Department of Physics, Naresuan University, Phitsanulok 65000, Thailand
}

\author{Natthaphat Thongyoo}
\author{Sukanya Sombun}
\author{Ayut Limphirat} 
\author{Yu-Peng Yan}
 \email{Contact corresponding author : yupeng@g.sut.ac.th}
\affiliation{
 School of Physics and Center of Excellence in High Energy Physics and Astrophysics, Suranaree University of Technology, Nakhon Ratchasima 30000,Thailand
}

\date{\today}

\begin{abstract}
We investigate the production of the scalar meson $f_0(980)$ in proton--proton collisions at $\sqrt{s}=5.02$~TeV using the Ultra-relativistic Quantum Molecular Dynamics (UrQMD) transport model supplemented with a $K\bar{K}$ coalescence afterburner. After conservatively tuning the UrQMD string-fragmentation parameters, the model reproduces the bulk charged-kaon production in the low-to-intermediate transverse-momentum region, providing the kaon phase-space distribution used as input for the coalescence calculation. In the present implementation, both charged and neutral kaon--antikaon pairs are considered, and each accepted $K\bar{K}$ pair is assigned to the isoscalar $f_0(980)$ and isovector $a_0(980)$ channels with an equal Monte Carlo probability. Using the updated integration analysis, we find that $\Delta p=0.4$~GeV/$c$ gives the best directly simulated agreement with the ALICE $p_T$ spectrum and integrated yield, while a linear interpolation between the neighboring points at $\Delta p=0.3$ and $0.4$~GeV/$c$ yields an interpolated optimum of $\Delta p^{\ast}\approx0.365$~GeV/$c$. Within this constrained hadronic coalescence framework, the measured $f_0(980)$ production is reasonably described, and the results are consistent with interpreting the $f_0(980)$ as a late-stage $K\bar{K}$ molecular configuration formed near kinetic freeze-out in small collision systems.
\end{abstract}

\maketitle

\section{Introduction}

Quantum chromodynamics (QCD), the fundamental theory of strong interaction, consistently describes interactions among quarks and gluons and elucidates color confinement. This theory clarifies the mechanisms by which quarks and gluons come together to make ordinary hadrons such as mesons (which consist of a quark and an antiquark ($q\bar{q}$)) or baryons (which contain three quarks ($qqq$)). In addition to these conventional configurations, QCD also allows a whole class of exotic hadronic states like tetraquarks ($q\bar{q}q\bar{q}$), pentaquarks  ($qqqq\bar{q}$) and multi-gluon bound states \cite{PhysRevD.15.267,Z.PhysA.359.305.1997,PhysRevD.60.034509}. These systems continue to be at the cutting edge of theory and experiment. If there are also excitations of the quarks responsible for forming bounding and scattering states, you could be sure that some part of these baryons did not resemble the expectations that one gets from a consideration with just pions or standard QCD considerations \cite{10.1093/ptep/ptaa104}. Importantly, the first observation of the $X(3872)$ at Belle offered encouraging evidence for tetraquark candidates \cite{PhysRevD.97.054513} and there are recent findings regarding pentaquarks from the LHCb collaboration \cite{Aaij2022}.

Among these exotic candidates, the $f_0(980)$ particle represents a particularly intriguing case due to its disputed internal structure. First observed through its decay into pion-pion states during the 1970s \cite{Fachini_2004,PhysRevD.7.1279,HYAMS1973134}, this resonance has sparked considerable theoretical debate. Current interpretations suggest three primary structural possibilities: a conventional quark-antiquark meson ($q\bar{q}$) \cite{PhysRevD.67.094011}, a compact four-quark state ($qq\bar{q}\bar{q}$) \cite{PhysRevD.15.267,PhysRevD.15.281}, or a $K\bar{K}$ molecular bound state \cite{PhysRevD.67.094011,PhysRevD.27.588}. Previous studies on the $f_0(980)$ meson have characterized it as a quarkonium state, representing a bound state of $q\bar{q}$. It has been shown that the $f_0(980)$ particle contains an $s\bar{s}$ component mixed with a non-strange quark-antiquark pair $\left (  u\bar{u}+d\bar{d}  \right )/\sqrt{2}$, with $s\bar{s}$ accounting for 10\% of the entire state \cite{efimov1993quark,tornqvist1995understanding,van1999comment,PhysRevC.61.035201,anisovich2003quark,anisovich2005riddlef0980a0980quarkantiquark}. The $f_0(980)$ meson has been identified as the lightest cryptoexotic $q\bar{q}$ nonet\cite{PhysRevD.15.267}. Progress has also been made in understanding the four-quark model, as documented in references \cite{ACHASOV1982134,PhysRevD.74.014028}. Additionally, studies on $\phi$ decay have provided insights into the structure of $f_0(980)$. However, the analysis highlights the challenges in distinguishing whether $f_0(980)$ is a $q\bar{q}$ meson or a four-quark ($qq\bar{q}\bar{q}$) state \cite{10.1143/PTPS.168.173}. The nature of $f_0(980)$ remains a subject of ongoing debate, with differing conclusions in the literature about its unique structure. Recently, the CMS experiment at CERN has provided strong evidence supporting the hypothesis that $f_0(980)$ is an ordinary quark-antiquark meson, based on measurements of its elliptic anisotropy in proton-lead collisions at 8.16 TeV \cite{cmscollaboration2023ellipticanisotropymeasurementf0980, 10.1093/ptep/ptaa104}. This finding adds to the complexity of understanding $f_0(980)$’s structure and production mechanisms. However, these experimental findings have not definitively resolved the structural ambiguity of $f_0(980)$. Alternative interpretations, such as the $K\bar{K}$ molecular picture or compact tetraquark configurations, continue to warrant theoretical and experimental investigation using complementary methodologies. One prevalent hypothesis suggests that the $f_0(980)$ meson is primarily a $q\bar{q}$ state, but its structure is significantly influenced by its proximity to the $K\bar{K}$ threshold. This proximity enhances its interaction with the kaon-antikaon pair, adding a substantial $K\bar{K}$ component to its makeup. Consequently, the $f_0(980)$ exhibits properties that arise from a combination of its $q\bar{q}$ core and the effects of the $K\bar{K}$ interaction, resulting in a complex structure that may include both $q\bar{q}$ and $K\bar{K}$ virtual components \cite{tornqvist1995understanding}. In reference \cite{Branz_2008}, the $f_0(980)$ meson is studied as a $K\bar{K}$ molecule using a phenomenological Lagrangian approach, which calculates its decay width and coupling constants with $K$ and $\bar{K}$. This analysis supports the hypothesis of its molecular nature. The $f_0(980)$ meson, often observed in scattering experiments, is regarded as a resonance originating from a $K\bar{K}$ molecule \cite{PhysRevD.97.054513,PhysRevD.101.094034}. The mass of $f_0(980)$ is close to the $K\bar{K}$ threshold, which results in strong coupling to $K\bar{K}$. This strong coupling facilitates its detection in experiments, making it more observable compared to hypothetical tetraquark states ($qq\bar{q}\bar{q}$) \cite{PhysRevD.101.094034}. Tetraquark states are structurally more complex and less understood than meson resonances like $f_0(980)$ because tetraquarks involve four quarks (two quarks and two antiquarks) bound together, which leads to intricate interactions between the quarks. Unlike mesons, which are composed of a single quark-antiquark pair, tetraquarks require the study of multiple potential configurations, such as diquark-antidiquark structures or molecular-like states. Furthermore, the dynamics of how these quarks bind, their decay channels, and their experimental signatures remain challenging to model and detect. Scattering experiments face additional difficulties in producing tetraquarks due to these complexities, requiring specific experimental conditions for their detection \cite{PhysRevD.101.094034}. In contrast, $f_0(980)$ is easier to observe due to its strong coupling to $K\bar{K}$ and its well-defined resonance behavior in scattering experiments \cite{PhysRevD.97.054513,PhysRevD.101.094034}.

Given the persistent ambiguities surrounding the internal structure of the $f_0(980)$, theoretical investigations based on dynamical transport models remain essential for clarifying its production mechanism in high-energy collisions~\cite{Cho2017ExoticHadrons,AhmedXiao2020}. This study investigates $f_0(980)$ production in proton-proton ($pp$) collisions at $\sqrt{s}=5.02$ TeV within the Ultra-relativistic Quantum Molecular Dynamics (UrQMD) model~\cite{Bleicher1999UrQMD,Bass1998MicroscopicModels}. As a first step, the UrQMD output is benchmarked against ALICE data for charged-kaon production at mid-rapidity $ \left | y \right |$ $<$ 0.5 over 
the transverse-momentum range 0 $<$ $p_{T}$ $<$ 10 GeV/c,as well as the integrated kaon yields, thereby establishing a reliable phase-space baseline for the subsequent coalescence analysis~\cite{Acharya2020Kaons}. To model $f_0(980)$ formation, we implement a coalescence afterburner applied to the UrQMD output, in which kaon-antikaon pairs are tested using relative-momentum ($\Delta p$) and spatial-separation ($\Delta r$) criteria at kinetic freeze-out~\cite{GuWang2023,Hillmann2021Coalescence,Sombun2018Deuteron,Li2016Helium3}. In the updated treatment, both the charged ($K^{+}K^{-}$) and neutral ($K^{0}\bar{K}^{0}$) channels are explicitly included in the coalescence analysis. Since both channels contain components that can project onto the isoscalar $f_0(980)$ state in the $K\bar{K}$ system, their contributions must be incorporated consistently in the coalescence framework~\cite{Achasov2019IsospinBreaking,Workman2022PDG}. To determine suitable values of the coalescence parameters, we perform a scan of the relative momentum cut-off over the range $0.1 \le \Delta p \le 1.0$ GeV/$c$, while requiring $0 \le \Delta r \le 3.0$ fm, consistent with earlier coalescence studies~\cite{GuWang2023,Hillmann2021Coalescence,Sombun2018Deuteron,Li2016Helium3}. The resulting transverse-momentum spectra and integrated yields of $f_0(980)$ in $pp$ collisions are then compared directly with the corresponding ALICE measurements~\cite{Acharya2023f0}. A detailed discussion of the coalescence setup, parameter scan, and comparison with experiment is presented in Sec.~III.

\section{\texorpdfstring{$f_{0}(980)$}{f0(980)} in Molecular Picture in UrQMD}
\label{sec:II}

UrQMD (Ultra-relativistic Quantum Molecular Dynamics) is a microscopic hadronic transport model widely used to simulate high-energy hadronic and nuclear collisions~\cite{Bleicher1999UrQMD,Bass1998MicroscopicModels}. It provides the space-time and momentum distributions of final-state hadrons after hadronic rescatterings and resonance decays, and at sufficiently high collision energies it also incorporates the excitation and fragmentation of color strings into hadrons~\cite{Bleicher1999UrQMD}. In the present study, UrQMD is employed to generate the phase-space distributions of final-state kaons and antikaons that serve as the input for the subsequent coalescence analysis.

Within the molecular picture, the $f_{0}(980)$ is treated as a late-stage $K\bar{K}$ bound configuration formed near kinetic freeze-out~\cite{Branz_2008,AhmedXiao2020}. Since the standard UrQMD particle content does not explicitly contain a production channel for a bound $K\bar{K}$ $f_{0}(980)$ state, the resonance is introduced through an external coalescence afterburner applied to the final-state kaon ensemble~\cite{Reichert2024,GuWang2023}. This approach is analogous to coalescence treatments widely used for the formation of composite hadrons and light nuclei in transport-model studies~\cite{Hillmann2021Coalescence,Sombun2018Deuteron,Li2016Helium3}.

More specifically, after the UrQMD evolution has concluded, the coalescence afterburner explicitly examines both charged kaon pairs, $K^{+}K^{-}$, and neutral kaon pairs, $K^{0}\bar{K}^{0}$, in the final state. A kaon-antikaon pair is accepted as a molecular candidate only if it satisfies the prescribed phase-space proximity conditions in relative momentum and relative distance. This updated treatment allows the contributions from both charged and neutral $K\bar{K}$ channels to be incorporated consistently in the construction of the final $f_{0}(980)$ yield.

To justify the isospin treatment implemented in the coalescence afterburner, one needs to consider the Clebsch--Gordan decomposition of the $K\bar{K}$ system. 
As the physical charged and neutral kaon-antikaon states can be decomposed as
\[
|K^{+}K^{-}\rangle
=
\frac{1}{\sqrt{2}}\,|I=1,I_{z}=0\rangle
+
\frac{1}{\sqrt{2}}\,|I=0,I_{z}=0\rangle,
\]
\[
|K^{0}\bar{K}^{0}\rangle
=
-\frac{1}{\sqrt{2}}\,|I=1,I_{z}=0\rangle
+
\frac{1}{\sqrt{2}}\,|I=0,I_{z}=0\rangle.
\]
both the $K^{+}K^{-}$ and $K^{0}\bar{K}^{0}$ contain the isoscalar and isovector components with equal weight~\cite{Achasov2019IsospinBreaking,Workman2022PDG}. Therefore, once a kaon-antikaon pair satisfies the phase-space coalescence criteria, the projection onto the isoscalar $f_{0}(980)$ channel and the isovector $a_{0}(980)$ channel is naturally implemented with equal probability, namely 50\% for $f_{0}(980)$ and 50\% for $a_{0}(980)$~\cite{Achasov2019IsospinBreaking}. 

In this framework, UrQMD supplies the realistic final-state kaon background, whereas the $f_{0}(980)$ signal is generated entirely through the external $K\bar{K}$ coalescence procedure. This separation is essential because it allows the molecular interpretation of the $f_{0}(980)$ to be tested directly against experimental data using the simulated kaon phase-space distributions as input.

\section{RESULTS AND DISCUSSION}

\subsection{Charged-Kaon Production in \emph{pp} Collisions}

\begin{table}[htb]
  \caption{Comparison of experimental data and UrQMD simulation results
  for the yield of charged kaons in \emph{pp} collisions at
  $\sqrt{s}=5.02\,$TeV. The experimental result includes systematic
  uncertainties.}
  \label{tab:comparison_yields}
  \begin{ruledtabular}
      \begin{tabular}{lcc}
      Particles & Experimental yield & UrQMD yield \\
      $K^+ + K^-$ & $0.534 \pm 0.027\,\text{sys.}$ & $0.527$ \\
      \end{tabular}
  \end{ruledtabular}
\end{table}

Table~\ref{tab:comparison_yields} shows that the tuned UrQMD 
calculation reproduces the integrated charged-kaon yield in 
inelastic $pp$ collisions at $\sqrt{s}=5.02$~TeV rather well. 
The ALICE measurement gives a mid-rapidity yield of 
$0.534 \pm 0.027$ sys., whereas the UrQMD result is $0.527$, 
in good agreement with the experimental 
value~\cite{Acharya2020}. This agreement provides an important 
benchmark for the kaon phase-space distributions used as input 
to the subsequent $K\bar{K}$ coalescence analysis.

To achieve this level of agreement, the UrQMD string-fragmentation parameters were conservatively tuned following the strategy motivated by Uzhinsky's studies of hadronic generator limitations at high energies~\cite{Uzhinsky:2011,Uzhinsky:2013}. Although hadronic transport models at LHC energies do not explicitly include hard partonic scatterings, the integrated kaon yield is dominated by the low-to-intermediate $p_T$ region, where string fragmentation governs particle production~\cite{Bleicher_1999,BASS1998255}. Therefore, a satisfactory description of the bulk kaon yield is sufficient for establishing a reliable baseline for the coalescence calculation.

\begin{figure}[htb]
  \centering
  \includegraphics[width=0.8\columnwidth]{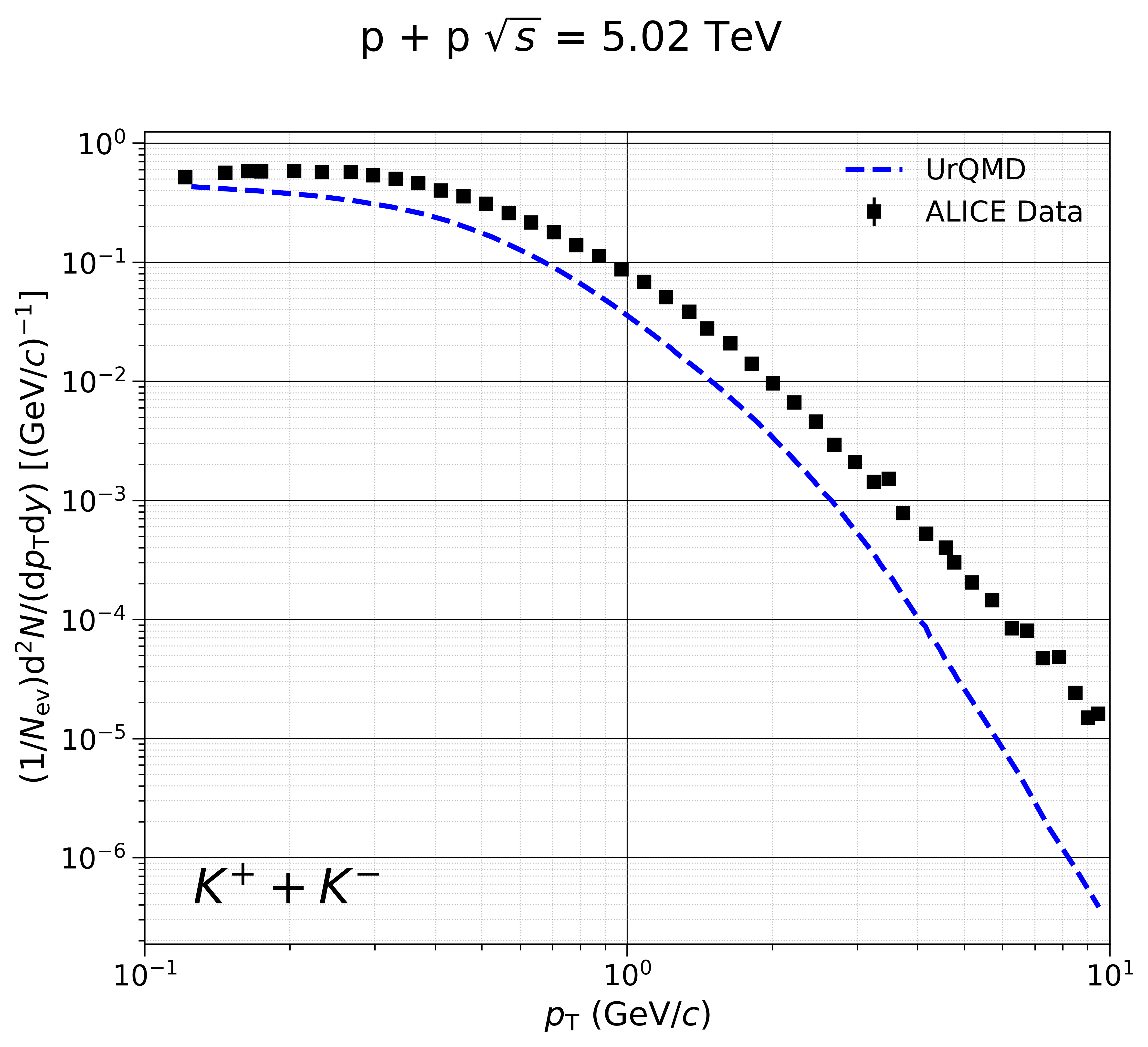}
  \caption{\label{fig:pt_distribution}Transverse-momentum spectra of charged kaons, $K^+ + K^-$, in proton--proton collisions at  $\sqrt{s}=5.02$~TeV within $|y|<0.5$. Black squares represent the ALICE data from Ref.~\cite{Acharya2020}, and the blue dashed line denotes the tuned UrQMD result.}
\end{figure}

Figure~\ref{fig:pt_distribution} compares the transverse-momentum spectrum of charged kaons from ALICE with the tuned UrQMD calculation. The model reproduces the shape and magnitude of the low-to-intermediate $p_T$ region reasonably well, while the discrepancy becomes more pronounced at higher $p_T$, where UrQMD underpredicts the data. This behavior is consistent with the known limitation of hadronic transport models, which do not contain the hard partonic processes needed to describe the high-$p_T$ tail accurately~\cite{Bleicher_1999,BASS1998255}.

Despite this high-$p_T$ deficit, the tuned UrQMD setup provides a reliable description of the bulk kaon production that dominates the coalescence probability. Since hadronic-molecule formation is strongly favored for constituents with small relative momentum and sufficiently small spatial separation, the low-to-intermediate $p_T$ kaons are the most relevant input to the present $K\bar{K}$ coalescence calculation~\cite{CHO2017279,Gu_2023}. The benchmark in Fig.~\ref{fig:pt_distribution} therefore supports the use of the UrQMD kaon ensemble as the baseline for constructing the $f_0(980)$ signal.

\subsection{Production of \texorpdfstring{$f_0(980)$}{f0(980)} in Proton--Proton Collisions}
Table II shows the integrated yields $\mathrm{d}N/\mathrm{d}y$ of $f_0(980)$ in  \emph{pp} collisions at $\sqrt{s}=5.02\,$TeV within $|y|<0.5$ derived in the UrQMD\,+\,coalescence model for various relative-momentum $\Delta p$. The target ALICE yield, $0.037887$, is obtained by Simpson integration of the experimental $p_T$ spectrum from Ref.~\cite{Acharya:2023plb846137644} using the same numerical procedure as applied to the model spectra. Linear interpolation between $\Delta p=0.3$ and 0.4\,GeV/c leads to an optimal value of $\Delta p = 0.365$\,GeV/c.

\begin{table}[htb]
	\caption{\label{tab:f0_yield_pp}
		Comparison of integrated yields $\mathrm{d}N/\mathrm{d}y$ of $f_0(980)$ in \emph{pp} collisions at $\sqrt{s}=5.02\,$TeV in the UrQMD\,+\,coalescence model with ALICE data. $\Delta$, as in the third column, gives the absolute difference between the model results and ALICE data. Linear interpolation between $\Delta p=0.3$ and $0.4$\,GeV/$c$ results in an optimal value of $\Delta p = 0.365$\,GeV/$c$.}
	\centering
	\begin{ruledtabular}
		\begin{tabular}{lccc}
			$\Delta p$ (GeV/$c$)& $\mathrm{d}N/\mathrm{d}y$ (UrQMD) & $\Delta$ & $R$ (UrQMD/ALICE)\\
					\hline
			0.1 & 0.000389 & 0.037498 & 0.0103 \\
			0.2 & 0.002967 & 0.034920 & 0.0783 \\
			0.3 & 0.032655 & 0.005233 & 0.8619 \\
			0.4 & 0.040724 & 0.002837 & 1.0749 \\
			0.5 & 0.050035 & 0.012148 & 1.3206 \\
			0.6 & 0.059597 & 0.021710 & 1.5730 \\
			0.8 & 0.077162 & 0.039275 & 2.0366 \\
			1.0 & 0.091292 & 0.053405 & 2.4096 \\
			\hline
			0.365 & 0.037887 & 0.000000 & 1.0000 \\
		\end{tabular}
	\end{ruledtabular}
\end{table}

\begin{figure}[!htb]
	\centering
	\includegraphics[width=0.82\columnwidth]{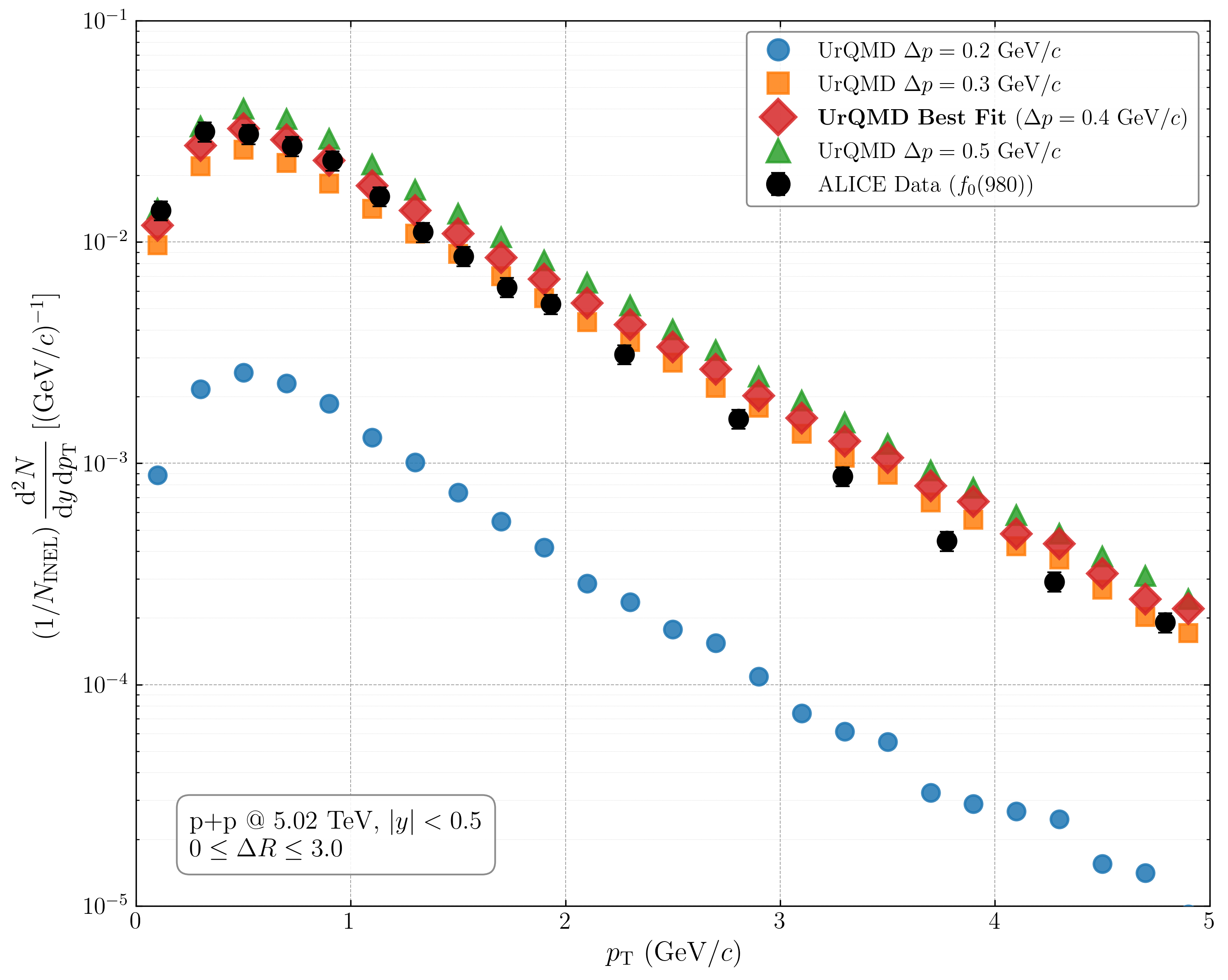}
	\caption{\label{fig:pt_distribution_f0_pp}Transverse-momentum spectra of $f_0(980)$ derived in $pp$ collisions at $\sqrt{s}=5.02$~TeV within $|y|<0.5$. Solid black circles represent the ALICE data from Ref.~\cite{Acharya:2023plb846137644}, while colored symbols stand for the UrQMD\,\,coalescence results for $\Delta p=0.2$, $0.3$, $0.4$, and $0.5$~GeV/$c$, respectively. The red diamonds highlight the best directly simulated results at $\Delta p=0.4$~GeV/$c$.}
\end{figure}

In Figure~\ref{fig:pt_distribution_f0_pp}, the UrQMD\,\,coalescence 
results of the transverse-momentum 
spectra of $f_0(980)$ in $pp$ collisions 
at $\sqrt{s}=5.02$~TeV within $|y|<0.5$ are compared with ALICE data. We show the results for only four selected values of the relative-momentum 
$\Delta p$,  $0.2$, $0.3$, $0.4$, and $0.5$~GeV/$c$. 
Among the directly simulated cases, the spectrum obtained with 
$\Delta p=0.4$~GeV/$c$ provides the best overall agreement with 
the ALICE data and is therefore identified as the best testing value.

In the present calculation, the coalescence afterburner examines 
both charged and neutral kaon--antikaon pairs that satisfy 
$\Delta p \leq p_{\max}$ and $0 \leq \Delta r \leq 3.0$~fm~\cite{Hillmann2021Coalescence,Sombun2018Deuteron,Li2016Helium3}. 
Following the isospin decomposition discussed in Sec.~\ref{sec:II}, 
each accepted $K\bar{K}$ pair is assigned to the isoscalar $f_0(980)$ 
channel or the isovector $a_0(980)$ channel with equal $50$--$50\%$ 
Monte Carlo probability. The integrated yields obtained from these 
spectra show that the target ALICE value is bracketed by the results 
at $\Delta p=0.3$ and $0.4$~GeV/$c$: the former slightly underpredicts 
the data, whereas the latter slightly overshoots it. A linear 
interpolation between these two neighboring points therefore gives 
an optimal cut-off of $\Delta p=0.365$~GeV/$c$.

In practical terms, $\Delta p=0.4$~GeV/$c$ is the best directly 
simulated choice and is thus identified as the best-fit curve in 
Fig.~\ref{fig:pt_distribution_f0_pp}, while $\Delta p^*=0.365$~GeV/$c$ 
represents the interpolated optimum inferred from the 
integrated-yield analysis. The overall agreement obtained in the 
vicinity of $\Delta p\approx 0.365$--$0.4$~GeV/$c$ supports the 
interpretation that the observed $f_0(980)$ yield can be described 
consistently within a late-stage $K\bar{K}$ coalescence picture 
near kinetic freeze-out.

\section{Summary}

In this study, we employed the Ultra-relativistic Quantum Molecular Dynamics (UrQMD) model~\cite{BASS1998255,Bleicher_1999}, complemented by a $K\bar{K}$ coalescence afterburner, to investigate the production of the scalar resonance $f_0(980)$ in proton--proton (pp) collisions at $\sqrt{s}=5.02$~TeV. Experimental measurements from the ALICE Collaboration~\cite{Acharya2020,Acharya:2023plb846137644} served as essential benchmarks. By conservatively tuning the UrQMD string-fragmentation parameters, our simulations successfully reproduced the bulk production of charged kaons in the low-to-intermediate transverse-momentum region, thereby providing a realistic and physically reliable kaon phase-space distribution for the subsequent coalescence study.

The key merit of this work is that the measured $f_0(980)$ transverse-momentum spectrum and integrated yield are well described within a constrained hadronic coalescence framework. In the present implementation, the $K\bar{K}$ coalescence afterburner examines both the charged and neutral kaon--antikaon pairs that satisfy the phase-space conditions in relative momentum and relative distance. Following the isospin decomposition discussed in Sec.~II, each accepted pair is assigned to the isoscalar $f_0(980)$ channel or the isovector $a_0(980)$ channel with equal Monte Carlo probability. Using the updated integration analysis, we find that $\Delta p=0.4$~GeV/$c$ provides the best directly simulated agreement with the ALICE data, while a linear interpolation between the neighboring points at $\Delta p=0.3$ and $0.4$~GeV/$c$ gives an interpolated optimum of $\Delta p= 0.365$~GeV/$c$.

Overall, the tuned UrQMD background combined with the constrained $K\bar{K}$ coalescence scheme provides a quantitatively reasonable description of $f_0(980)$ production in small collision systems at LHC energies. Within the present framework, the agreement of the theoretical results with experimental data strongly support the argument that the $f_0(980)$ interpreted as a late-stage $K\bar{K}$ molecular configuration is formed near kinetic freeze-out~\cite{Branz_2008,AhmedXiao2020,Hillmann2021Coalescence,Sombun2018Deuteron,Li2016Helium3}.

\begin{acknowledgments}
This research has received funding support from the NSRF via the Program Management Unit for Human Resources \& Institutional Development, Research and Innovation (PMU-B) [grant number B39G680010]. 
The authors gratefully acknowledge the Department of Physics, Faculty of Science, Naresuan University; the School of Physics and Center of Excellence in High Energy Physics and Astrophysics, Suranaree University of Technology; and the Department of Physics, School of Science, University of Phayao, for providing the laboratory facilities and computational resources that made this research possible.  

We are indebted to \textbf{Prof.\ Marcus Bleicher} for valuable guidance on practical aspects of the UrQMD code and the implementation of the coalescence--afterburner mechanism. 

The authors acknowledge the use of OpenAI's \emph{ChatGPT} and Google's \emph{Gemini 3.1 Pro} models for assistance with language polishing and stylistic refinement. Additionally, \emph{Perplexity.ai} was utilized to support literature search and information retrieval. All scientific content, data analysis, and final conclusions remain entirely the responsibility of the authors.

\end{acknowledgments}

\bibliography{apssamp}

\end{document}